# A Refined Equilibrium Generative Adversarial Network for Retinal Vessel Segmentation

Yukun Zhou, Zailiang Chen, Hailan Shen*, Xianxian Zheng, Rongchang Zhao, Xuanchu Duan

*Abstract*—*Objective:* Recognizing retinal vessel abnormity is vital to early diagnosis of ophthalmological diseases and cardiovascular events. However, segmentation results are highly influenced by elusive vessels, especially in low-contrast background and lesion region. In this work, we present an end-to-end synthetic neural network, containing a symmetric equilibrium generative adversarial network (SEGAN), multi-scale features refine blocks (MSFRB), and attention mechanism (AM) to enhance the performance on vessel segmentation. *Method*: The proposed network is granted powerful multi-scale representation capability to extract detail information. First, SEGAN constructs a symmetric adversarial architecture, which forces generator to produce more realistic images with local details. Second, MSFRB are devised to prevent high-resolution features from being obscured, thereby merging multi-scale features better. Finally, the AM is employed to encourage the network to concentrate on discriminative features. *Results:* On public dataset DRIVE, STARE, CHASEDB1, and HRF, we evaluate our network quantitatively and compare it with state-of-the-art works. The ablation experiment shows that SEGAN, MSFRB, and AM both contribute to the desirable performance. *Conclusion:* The proposed network outperforms the mature methods and effectively functions in elusive vessels segmentation, achieving highest scores in Sensitivity, G-Mean, Precision, and F1-Score while maintaining the top level in other metrics. *Significance:* The appreciable performance and computational efficiency offer great potential in clinical retinal vessel segmentation application. Meanwhile, the network could be utilized to extract detail information in other biomedical issues.

## I. Introduction

Retinal vessel morphology is valuable indicator for ophthalmological and cardiovascular diseases, such as diabetes, hypertension, and arteriosclerosis [1], [2]. Retinal vessel segmentation provides various morphological vessel features, which may provide a reliable reference required for quantitative analysis of such diseases [3]. However, manual segmentation only by Human observer is tedious and time-consuming. In this case, automatic vessel segmentation plays an increasingly important role in disease recognition and prevention [4]. In recent years, lots of researchers conduct novel work on improving the performance on vessel segmentation. Basically, the related approaches could be divided into two categories, supervised and unsupervised.

Unsupervised methods, requiring no manual annotation, mainly include matched filtering, vessel tracking, morphological transformations, and model-based algorithms. Rangayyan *et al.* [5] present a vessel tracking method by employing a Gabor filters to extract the vessels. Mendonca *et al.* [6] detect vessel ridges with multiple structuring elements. Neto *et al.* [7] develop a course-to-fine algorithm, relying on the mathematical morphology, spatial dependency and curvature. Zhao *et al.* [8] present an infinite active contour model by using hybrid region information. Zhang *et al.* [9] use a matched filter with first-order derivative of a Gaussian filter to segment vessels. Ali-Diri *et al.* [10] make use of two pairs of contours to locate vessel edge. Fraz *et al.* [11] use the first-order derivative of Gaussian filter for centerlines extraction with a morphological operator for morphology calculation. Roychowdhury *et al.* [12] present an adaptive thresholding method to complete iteration vessel segmentation. Salazar-Gonzalez *et al.* [13] first carry out a pre-processing for the image by adaptive histogram equalization and robust distance transform. Yin *et al.* [14] propose a segmentation method using hessian matrix and thresholding entropy, using post-processing to eliminate noise and the central light reflex. Fathi and Ahmad [38] vessel propose an vessel enhancement method based on complex continuous wavelet transform. Zhang et al. [40] present a robust and fully automatic filter-based approach for retinal vessel segmentation.

Supervised segmentation methods utilize ground truth vessel data to train a classifier in discriminating whether a pixel is vessel or non-vessel. Specifically, certain approaches need handcrafted features for segmentation, including K-nearest neighbor (KNN) [15], support vector machine (SVM) [16] and others. Ricci and Perfetti [17] employ line operators as feature vectors and SVM for pixel classification. Fraz *et al.* [18] use an ensemble classifier of boosted and bagged decision trees to construct supervised method for retinal image analysis. Roychowdhury *et al.* [19] reduce the pixels under classification by eliminating the major vessels that are detected as regions common to threshold versions of high-pass filtered images to save time. Lupascu *et al.* [20] employ different scale filters to extract 41D features for encoding information on local intensity structure, spatial properties, and geometry. With the rapid development of deep neural network, Li *et al.* [21] propose a wide and deep neural network that needs no artificially designed fea-

Manuscript submitted Nov 5, 2019. This work was supported by the National Natural Science Foundation of China under Grant No. 61672542 and 61972419. *Asterisk indicates corresponding author.*

H. Shen, Z. Chen, X. Zheng, and R. Zhao are with the Department of Computer Science and Technology, Central South University, Changsha, Hunan, 410083 China. The corresponding author is H. Shen (e-mail: hn_shl@126.com).
Y. Zhou was with Beihang University, Beijing, 100191 China. He is now with Department of Computer Science and Technology, Central South University, Changsha, Hunan, 410083 China.
X. Duan is with Changsha Aier Eye Hospital, Changsha, Hunan, 410015, China.

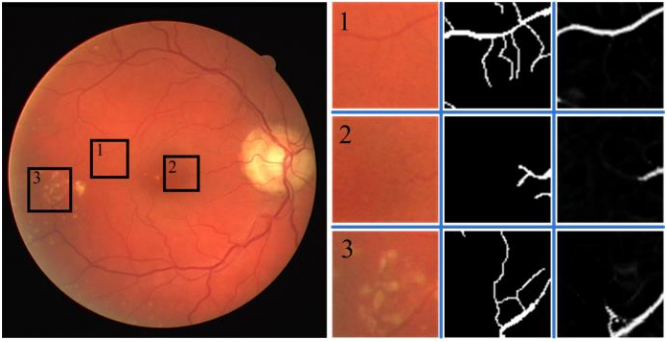

Fig. 1. The retinal fundus vessel segmentation results with U-Net [25]. From left to right, the four columns are fundus image from DRIVE, local patches in challenging situation, ground truth, and segmentation image with U-Net. According to local observation, the detail vessel information is lost, especially in regions with exudates (3), low contrast (1) and (2).

ture and preprocessing step, thereby reducing the impact of subjective factors. Wu et al. [22] present a multi-scale network followed network model to help segment the blood vessels, particularly the capillaries. Orlando et al. [23] put forward a discriminatively trained segmentation model on the base of fully connected conditional random fields, this model better distinguish the desired structures than the local neighborhood-based approach. Yan et al. [24] propose a new segment-level loss that emphasizes the thickness consistency of thin vessels in the training process, when considering highly imbalanced pixel ratios between thick and thin vessels in fundus images. Wang et al. [42] propose two encoders to preserve the spatial information and semantic information, and introduce a feature fusion module for vessel segmentation. Lyu et al. [39] propose a method utilizing separable spatial and channel flow and densely adjacent vessel prediction to capture maximum spatial correlations between vessels. Wu et al. [43] design an efficient inception residual convolutional block and introduce four supervision paths to preserve the multi-scale features.

With these studies on the vessel segmentation, the performance of vessel pixel classification has been increasingly improved, and the majority of vessels are able to be recognized. Several metrics such as the *accuracy* (*Acc*), *specificity* (*Sp*) and *area under curve* (*AUC*) are considerably increased. However, *Sensitivity* (*Se*), which is the proportion of actual vessel pixels that are correctly identified, is relatively low. Considering the definition of *Se*, the ratio of true vessel being detected is limited in low level. This situation results from the imbalanced distribution in a retinal fundus image. In recent work, even if some vessels are ignored in segmentation, the *Acc* and *Sp* still obtain high score as there are much more non-vessel pixels in the fundus image. Concentrating only on the *Sp* and *Acc* would lead to much loss during elusive vessels extraction. Thus, we need to achieve a trade-off between *Se*, *Sp*, and other comprehensive metrics, such as *AUC*, *G-Mean*, and *F1-Score*. By reviewing the previous research, there are few algorithms that achieve satisfied score on *Se*, which shows the elusive vessel is the huge obstacle influencing the performance of retinal vessel segmentation. As shown in Fig. 1, the U-Net developed by Ronneberger et al. [25] shows weak performance on the elusive vessels, especially in complex environment backgrounds, such as lesion and low-contrast surroundings. Elusive thin vessel pattern is significant in disease analysis, for instance, the first manifestations of diabetic retinopathy (DR) include tiny vessel dilations, known as microaneurysms (MA) and exudates [26]. Such manifestations may provide an early indication of the risk of the type I diabetes.

For further improving vessel segmentation effects, especially on elusive vessel recognition, we firstly propose Symmetric Equilibrium Generative Adversarial Networks (SEGAN), which utilizes the characteristics of U-Net and Generative Adversarial Network (GAN) [27]. In contrast to the conventional GAN, which only uses advanced structures as generator (G), such a VGG [28], Res-Net [29], U-Net, and Google-Net [30], we construct symmetric equilibrium architecture by employing U-Net as baseline both in the G and discriminator (D). Specifically, this structure eliminates the imbalance in inborn capability between G and D. The D shares the same U-Net structure with G so that they are in well-matched game. Second, we present the multi-scale feature refine blocks (MSFRB) to optimally merge the different scale features. MSFRB preserves high-resolution features with high-semantic ones simultaneously, aiming at keeping the multi-scale representation independent and refining much better local detail information. Finally, the attention mechanism (AM) is employed. By distributing larger weights, it highlights the discriminative feature maps rather than the inconsequential ones. In this case, the distinguishable features could be further strengthened.

Basically, this paper presents three contributions on the retinal vessel segmentation based on the present research.

1. We propose SEGAN for precise retinal vessel segmentation by utilizing adversarial principle to strengthen the G (i.e., U-net) capability. In addition, we build a symmetric adversarial architecture which allows D thoroughly distinguishes detail difference between output of G and ground truth, thereby forcing G to fake the details perfectly and enhancing the recognition ability for elusive vessels.
2. The MSFRB are presented to fully utilize the shallow-layer features which are high resolution but low semantic. In combination with deep-layer features. The multi-scale information is effectively merged, which avoids convolution confusion occurred in the traditional skip connection. The lightweight structure ensures the high efficiency for retinal vessel segmentation.
3. AM is employed in the MSFRB to train the network in allocating weights to different channels and concentrating on informative feature maps while ignoring valueless ones. Additionally, two extra weighted segmentation loss functions, namely, *binary cross-entropy loss* (*BCE*) and *mean absolute error* (*MAE*), have been included besides the conventional GAN loss function. It constructs an optimized objective function to spare more attention on pixel-level segmentation task.

The rest of this paper is organized as follows. We detail elaborate on the proposed method in Section II. In Section III, we describe the datasets used in the experiment and evaluation metrics. In Section IV, we present the abundant experiment results and compare our approach with other state-of-the-art methods. We then analyze the importance of SEGAN, MFSRB and AM by the ablation studies. Discussion is illustrated in Section V. Finally, we conclude this paper in Section VI.

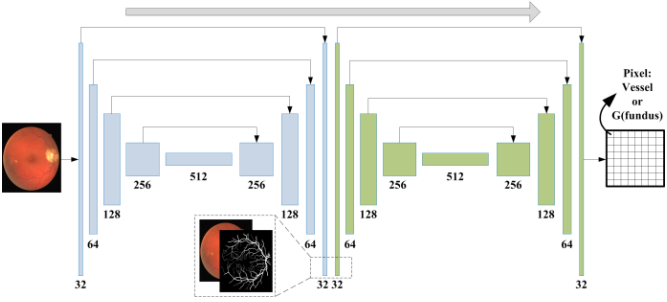

Fig. 2. Structure of SEGAN. Blue part represents G and green part indicates D. The running schedule is from left to right, indicated by top gray arrow. The black arrows in the G and D are skip-connections used for multi-scale features fusion. Each block denotes a stage and the channel number is listed below the block. The output of D is a map with the same size of retinal image, and value of each pixel represents the possibility of being ground truth *vessel*.

## II. METHODS

### A. Symmetric Equilibrium Architecture

In vessel segmentation, adversarial training in GAN [27] could be utilized to improve the G capability. Let $F_G$ refers to the function from fundus image *fundus* to vessel image *vessel*, and $F_D$ is function from (*fundus*, *vessel*) to binary classification (0, 1). Then the conventional loss function is defined as follow.

$$\mathcal{L}_{GAN} = \min_{\theta_G} \max_{\theta_D} \mathbb{E}_{fundus, vessel \sim p_{data}(f,v)}[\log F_D(fundus, vessel)] \\ + \mathbb{E}_{fundus \sim p_{data}(f)}[\log(1 - F_D(fundus, F_G(fundus)))] \quad (1)$$

where $\theta_G$ and $\theta_D$ represent the parameters in the G and D respectively. The D is trained to maximize the objective function (i.e., $F_D(fundus, vessel) \approx 1$ and $F_D(fundus, F_G(fundus)) \approx 0$). By contrast, G is trained to minimize the objective function, that is, $F_G(fundus)$ is extremely indistinguishable with *vessel*.

The motivation is that, in traditional methods, some networks such as U-Net [25], are employed in G to achieve $F_G(fundus) \approx vessel$. U-Net is a powerful feature extraction network, especially in biomedical image analysis. However, research on the D structure is scarce. Supposing that discriminator is weak, $F_D(fundus, F_G(fundus))$ will approximate a value of 1, even if $F_G(fundus)$ is imperfect. In more specific, although D does not segment vessels directly, it requires a strong capability to recognize the detail difference between $F_G(fundus)$ and *vessel*, otherwise the $F_G(fundus)$ which losses lots of elusive vessels would be regarded as true *vessel*. In this case, the training on the G gains no ideal result with the weak adversarial environment.

In order to grant the G stronger capability to extract the high resolution information, we propose the symmetric equilibrium architecture by using U-Net as baseline both in G and D of GAN. The overall network of SEGAN is concisely shown in Fig. 2, without any pre-processing and post-processing. The two sides of the network are symmetric. The left side is G which takes the retinal fundus image as input and outputs the vessel probabilistic map of retinal vessels. The vessel probabilistic map is then concatenated with retinal image and delivered to the D for evaluation. The D contains five stages in the down-sampling process to obtain high semantic information. Each stage consists of two convolution layers, two batch normalizations, two activations and one max-pooling layer for deep feature extraction. After the five stages, the high semantic feature map proceeds to up-sampling to recover to the original size. Five stages that include 2x up-sampling also exist. In each stage, the feature map from shallow layers is concatenated with the up-sampling feature map through skip connection, combing the low semantic but high-resolution features with high semantic but low-resolution ones. Being different with the traditional D, the output of presented D is a possibility map which has the same size as retinal image, which means it discriminate the $F_G(fundus)$ and *vessel* in each pixel. This structure endows D the capability to recognize the detail difference. Accordingly, the D re-emphasizes the significance of not only thick vessel trunk, but also elusive vessels which is full of detail information. In this well-matched setting, adversarial training is strengthened to force the G segment more realistic vessel images (i.e., $F_G(fundus) \approx vessel$).

### B. Multi-Scale Features Refine Block

In traditional up-sampling process of G, the high resolution features from the shallow layers are concatenated with high semantic ones. The concatenated layers are then delivered to the next convolution, as shown in Fig. 3(a). The convolutional computation is efficient to extract discriminative features, but it sacrifices the high resolution details. Although the traditional multi-scale merging process recognizes local details to a certain degree, the high resolution features are confused with high semantic ones by convolution operation, lacking representation preservation. This phenomenon hinders the vivid reconstruction of elusive thin vessels. Consequently, we propose MSFRB to refine the two types of features and advance the merging process, as shown in Fig. 3b.

The features at stage $s$ in down-sampling and up-sampling are denoted by $x_s^d$ and $x_s^u$, $s \in \{1, 2, ..., N^{d\,or\,u}\}$. In U-Net, $N^d$ and $N^u$ are both equal to 5. The $x_{s-1}^u$ can be calculated as follow.

$$x_{s-1}^u = \text{MSFRB}(x_s^u, x_s^d, x_1^d) \quad (2)$$

where MSFRB() contains six steps. The inputs are high resolution features $x_s^d$, $x_1^d$ and high sematic feature $x_s^u$. First, it completes a convolution operation for the previous stage output $x_s^u$ and the first feature map $x_1^d$.

$$\tilde{x}_s^u = \text{Conv2d}(x_s^u, n_c(x_s^d)) \quad (3)$$

$$\tilde{x}_1^u = \text{Conv2d}(x_1^u, n_c(x_s^d), n_s(x_s^d)) \quad (4)$$

where Conv2d() represents 2D convolution operation, and $n_c(x_s^d)$, $n_s(x_s^d)$ are the parameter of Conv2d function, which defines the convolution kernel channel and size. $\tilde{x}_s^u$ and $\tilde{x}_1^d$ denote the result of convolution, which has the same size and

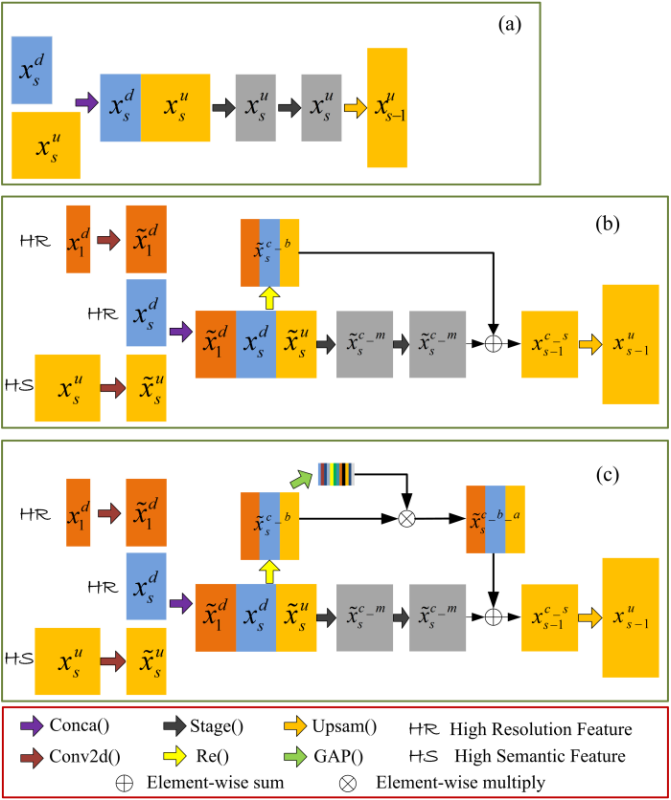

Fig. 3. The process graphs of the multi-scale features refine blocks (MSFRB). (a) Conventional fusion process, (b) Features fusion process with proposed MSFRB, (c) With MSFRB and AM module.

channel as $x_s^d$

Then, it concatenates $\tilde{x}_s^u$, $\tilde{x}_1^d$, and $x_s^d$ with function Conca(). This step combines high resolution features with high semantic ones.

$$\tilde{x}_s^c = \text{Conca}(\tilde{x}_1^d, \tilde{x}_s^u, x_s^d) \quad (5)$$

Subsequently, we divide the algorithm schedule into two parts: *main road* and *branch road*, as shown in Fig. 3(b). These two parts are separately defined as follows.

$$\tilde{x}_s^{c\_m} = \text{Stage}(\tilde{x}_s^{c\_m}) = \text{Iteration}_{n=2}(\text{ReLu}(\text{BN}(\text{Conv}2(\tilde{x}_s^{c\_m})))) \quad (6)$$

$$\tilde{x}_s^{c\_b} = \text{Re}(\tilde{x}_s^u, x_s^d) = \text{Conca}(r(\tilde{x}_s^u), r(x_s^d)) \quad (7)$$

where $\tilde{x}_s^{c\_m}$ represents the component of the main part and $\tilde{x}_s^{c\_b}$ notes the value of the branch part. The Stage() function expresses the operation conducted in the main road, which includes two iteration rounds containing Convolution Conv2(), BatchNormalization BN(), and ReLu Activation ReLu(). In the branch part, Re() is a significant function in maintaining high resolution and semantic feature representations and consists of channel sum function r() and Conca(). Letting $a=[a_1, a_2, ..., a_n]$, the definition of r() is as follow

$$r(a) = a' = [a'_1, a'_2, ...., a'_{n/k}]$$
$$a'_z = \sum_{i=1}^{k} a_{zk+i}, \ z \in [1, n/k] \quad (8)$$

In this function, the adjacent $k$ channels sum up to squeeze the channel, so that $\tilde{x}_s^{c\_b}$ has appropriate channel number to be merged with $\tilde{x}_s^{c\_m}$. The $\tilde{x}_s^{c\_b}$ avoids the convolution operation for the high resolution features, thus keeping the detail information from being obfuscated. Meanwhile, the $x_s^d$ and $\tilde{x}_s^u$, as inputs of Re(), are independently squeezed, which means the high resolution features and high semantic features are not confused together, occupying their room separately.

We sum up the main part element $\tilde{x}_s^{c\_m}$ and branch part element $\tilde{x}_s^{c\_b}$, and use the up-sample function Ups () to obtain the output of this stage.

$$x_{s-1}^{c\_s} = \tilde{x}_s^{c\_b} + \tilde{x}_s^{c\_m} \quad (9)$$

$$x_{s-1}^u = \text{Ups}(x_{s-1}^{c\_s}) \quad (10)$$

In MSFRB, the output feature map is composed of two parts, multi-scale features convolution fusion $\tilde{x}_s^{c\_m}$ and independent multi-scale features $\tilde{x}_s^{c\_b}$, which refine the segmentation performance in high resolution and semantic. Meanwhile, it introduces the highest resolution feature $x_1^d$ in the module so that the detail information could be supplemented in large degree.

### C. Attention Mechanism

Attention can be interpreted as a means of biasing the allocation of available computational resources toward the most informative components of a signal [31], [32].

During the MSFRB, some discriminative feature maps are added in, while a number of insignificant ones are also introduced. For highlighting the representation capability in vessel segmentation, the AM is utilized to focus on feature maps being beneficial to task.

As shown in Fig. 3(c), the attention module consists of global average pooling and activation. The application position is on the branch part in MSFRB, and the formula can be listed as follow.

$$L = \text{GAP}(\tilde{x}_s^{u\_b}) = [l_1, l_2, ..., l_N]$$
$$= [\sum_{i=1}^{w}\sum_{j=1}^{h} pixel_{1,i,j}, \sum_{i=1}^{w}\sum_{j=1}^{h} pixel_{2,i,j}, ..., \sum_{i=1}^{w}\sum_{j=1}^{h} pixel_{N,i,j}]/(w \cdot h) \quad (11)$$

$$Atten = \text{Sigmoid}(L) \quad (12)$$

where GAP() indicates the global average pooling function; $w$ and $h$ represent the width and height of image respectively, and *pixel* is the intensity of image pixel. The following activation Sigmoid() controls the attention matrix value at [0, 1], thereby avoiding gradient explosion in training process, simultaneously introducing non-linear representation in the attention module.

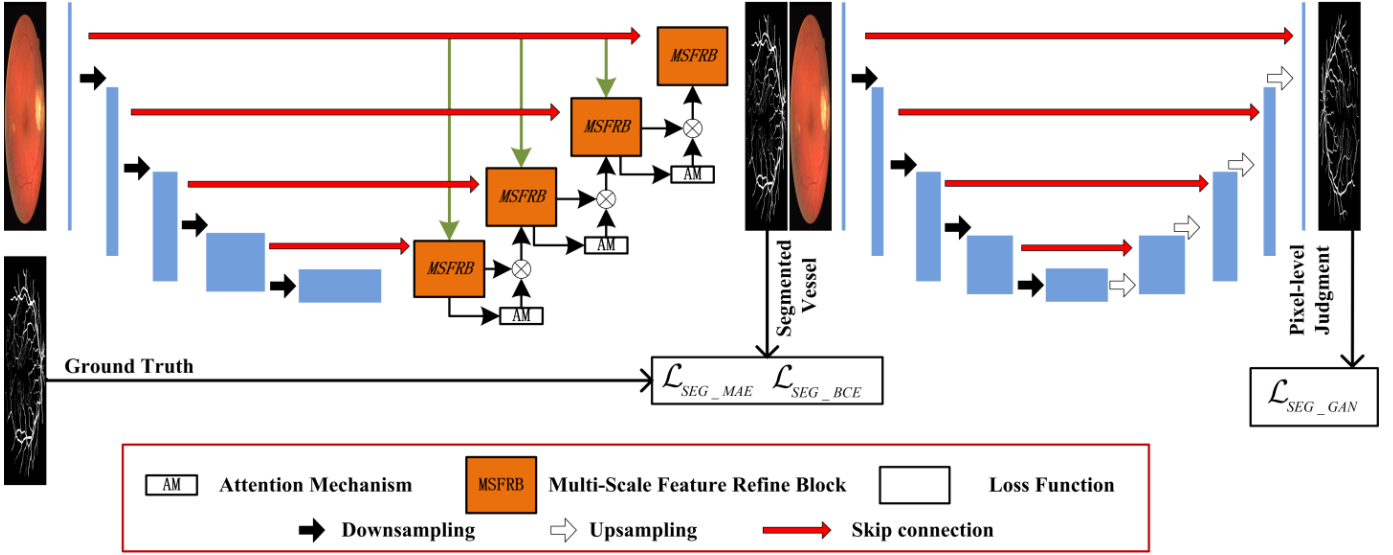

Fig. 4. The overall structure of proposed synthetic network, including SEGAN, MSFRB and AM.

The former value of branch part $\tilde{x}_s^{c\_b}$ is ameliorated by the *Atten*, which is computed by formula (11) and (12). And corresponding output features of stage s could be revised as follows.

$$\tilde{x}_s^{c\_b\_a} = Atten \cdot \tilde{x}_s^{c\_b} \quad (13)$$

$$x_{s-1}^u = \tilde{x}_s^{c\_b\_a} + \tilde{x}_s^{c\_m} \quad (14)$$

In comparison with the non-attention result, the attention matrix emphasizes the discriminative feature maps instead of dispersing concentration evenly on all channels. This case ensures that pixel classification behave much better with valuable information.

### D. Overall Network and Objective Loss Function

MSFRB together with AM are employed on the G, preserving multi-scale representation when allocating additional weights on informative feature channel, as shown in Fig. 4.

The objective loss function is vital to network training performance. Considering that our task is vessel segmentation (i.e., pixel classification), we define our objective loss function $\mathcal{L}$ as follows.

$$\mathcal{L} = \alpha \cdot \mathcal{L}_{GAN} + \beta \cdot \mathcal{L}_{SEG\_BCE} + \gamma \cdot \mathcal{L}_{SEG\_MAE} \quad (15)$$

where $\mathcal{L}$ contains three components, namely, GAN loss function $\mathcal{L}_{GAN}$, BCE $\mathcal{L}_{SEG\_BCE}$, and MAE $\mathcal{L}_{SEG\_MAE}$. $\alpha$, $\beta$, and $\gamma$ are hyperparameters used to allocate weights.

$\mathcal{L}_{SEG\_BCE}$ and $\mathcal{L}_{SEG\_MAE}$ are used to solve the tiny vessels detection challenge, as they directly evaluate the distance between ground truth and prediction. And their definition are briefly listed below.

$$\mathcal{L}_{SEG\_BCE} = \mathbb{E}_{x,y \sim p_{data(x,y)}} - y \bullet \log G(x) - (1-y) \bullet \log(1 - G(x)) \quad (16)$$

$$\mathcal{L}_{SEG\_MAE} = \mathbb{E}_{x,y \sim p_{data(x,y)}} |y - G(x)| \quad (17)$$

All the components are devised for enhancing the detail information learning capability of retinal vessel segmentation, so that the elusive vessel pixels, which are easily ignored by existing methods, are able to be recognized with the proposed symmetric network.

### III. MATERIALS AND EVALUATION

### A. Datasets

We complete experiments on the four public dataset DRIVE [7], STARE [33], CHASEDB1 [34], and HRF [35] to evaluate the retinal vessel segmentation effect. DRIVE dataset includes 40 color retinal fundus images with a resolution of (565, 584). Usually these images are divided into two parts. The first one is training set with 20 images containing one annotation per image, while the other one is test group which has 20 images with two labeled vessel images per image. One is used as ground truth and the other one as second human observer. STARE contains 20 retinal fundus images, and each image resolution is $700 \times 605$. Several studies used "leave one out" strategy to train 19 images and test on one image [6], [24]. We adopt the same strategy for fair comparison. CHASEDB1 dataset consists of 28 retinal fundus images with a resolution of $999 \times 960$ pixels. As normal, we divide it into a training set containing 20 images and a test set with 8 images. Additionally, the FOV mask in STARE and CHASEDB1 are not given in original set. So we separately use the STARE mask built in [36] and CHASEDB1 mask obtained in [23]. For HRF dataset, the resolution is $3504 \times 2336$. It has three subsets healthy, diabetic retinopathy, and glaucomatous, and each subset has 15 images. Like the [23] and [24], we employ the first 5 images in every subset as training dataset, and the left images as test dataset.

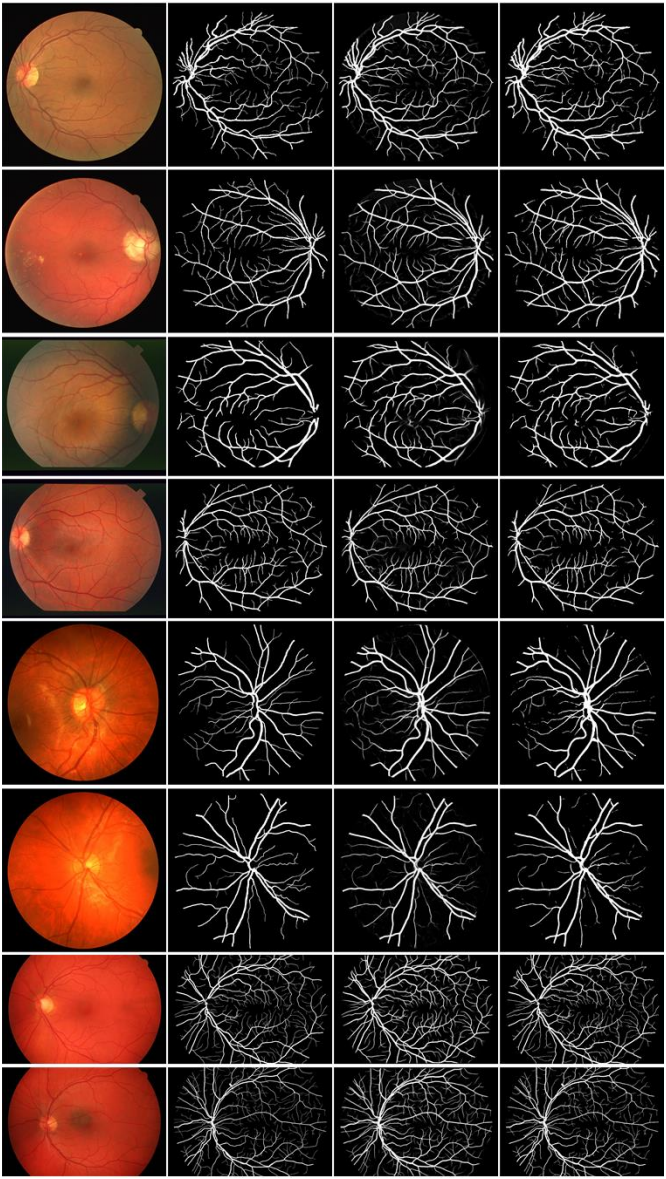

Fig. 5. Exemplar segmentation results on DRIVE (row 1, 2), STARE (row 3, 4), CHASEDB1 (row 5 to 6), and HRF (row 7, 8). The four columns from left to right: fundus images, ground truth, possibility map, and binary segmented vessel. The binary map is obtained with OSTU algorithm [41]. Better view with zoom in.

## B. Evaluation Metrics

We quantitatively analyze our experiment results and compared them with the label ground truth vessel. Based on the number of true positive *TP*, true negative *TN*, false positive *FP*, and false negative *FN*, seven metrics are employed.

The metrics computation formulas are as follows.

$$Se = \frac{TP}{TP+FN}, \quad Sp = \frac{TN}{TN+FP}, \quad Pr = \frac{TP}{TP+FP}$$
$$G = \sqrt{Se \times Sp}, \quad F1 = \frac{2 \times Pr \times Re}{Pr+Re} \quad (17)$$
$$Acc = \frac{TP+TN}{TP+FN+TN+FP}$$

*Se* represents the ability of correctly recognizing vessel pixels, which directly reflects the vessel segmentation ability. It is an appropriate metric to evaluate the improvement on elusive vessel pixels segmentation. *Sp* measures the capability on detecting non-vessel parts. *Precision* (*Pr*) indicates the proportion of pixels classified as vessels that are accurately identified. *Acc* is frequently employed to evaluate classifier performance. Additionally, we make use of two more indicators, *G-Mean* (*G*) and *F1-Score* (*F1*), for overall performance. *F1* is the harmonic mean of precision and recall, which owns the property of better characterizing quality when the data are imbalanced. *G* refers to the geometric mean of two most important metric *Se* and *Sp*, comprehensively estimating the pixel classification effects. The *receiving operator characteristic* (*ROC*) curve is computed with the *Se* versus (1−*Sp*) with respect to a varying threshold. The area under the ROC curve (*AUC*) is calculated for quality evaluation. All these metrics are equal to 1 under ideal conditions, while being 0 in worst classification.

## C. Implementation Details

In order to relax the computational stress and be convenient for subsequent cross-training experiments, we firstly down-sample the HRF and CHASEDB1 images (except for the ground truths in test dataset) using a factor of 4 and 2 respectively. Then we stipulate a fixed resolution (880, 608) for whole four datasets, except for the ground truths in test dataset. The training images are unified in size through padding operation. It is worth noting that the vessel images generated from test images are recovered to the original size and the padding zero are removed by mask, thus the evaluation results are fair to make comparison with the other methods. We believe that a large training image size is profitable for a network learning global semantic representation. Meanwhile, a large training image size avoids the problem of erroneous recognition of huge vessels as background in small patch training [22]. Data augmentation is supposed to expand the amount of images by rotation and flipping for network robustness and training performance. After augmentation, DRIVE and CHASEDB1 contains 2400 images in training set. STARE has 2280 images for training and HRF has 1800 images.

Apart from the initial normalization, no preprocessing or postprocessing is necessary. The entire training process is end-to-end. Learning rate is fixed at 0.0002, and the batch size is two. Fig. 2 shows the channel setting. The hyperparameters $\alpha$, $\beta$, and $\gamma$ are set at 0.08, 1.1, and 0.5 respectively. As the vessel segmentation is pixel classification task, we adopt 'Sigmoid' activation in the last output layer of G and D to obtain probability maps with a range of [0, 1]. In the other parts of network, activation functions are universally "ReLu", which provides fast converge efficiency without saturation area. The Adam with *beta1=0.5* is employed for the optimizer. We define one round as an iteration of all training images, and in practice we conduct 10 rounds and average the metric value of the final 5 rounds to achieve stable results.

## IV. EXPERIMENT RESULTS

### A. Vessel Segmentation

Intuitive vessel segmentation results are drawn in Fig. 5, including DRIVE, STARE, CHASEDB1, and HRF. The segmentation maps are of little difference with the ground truths.

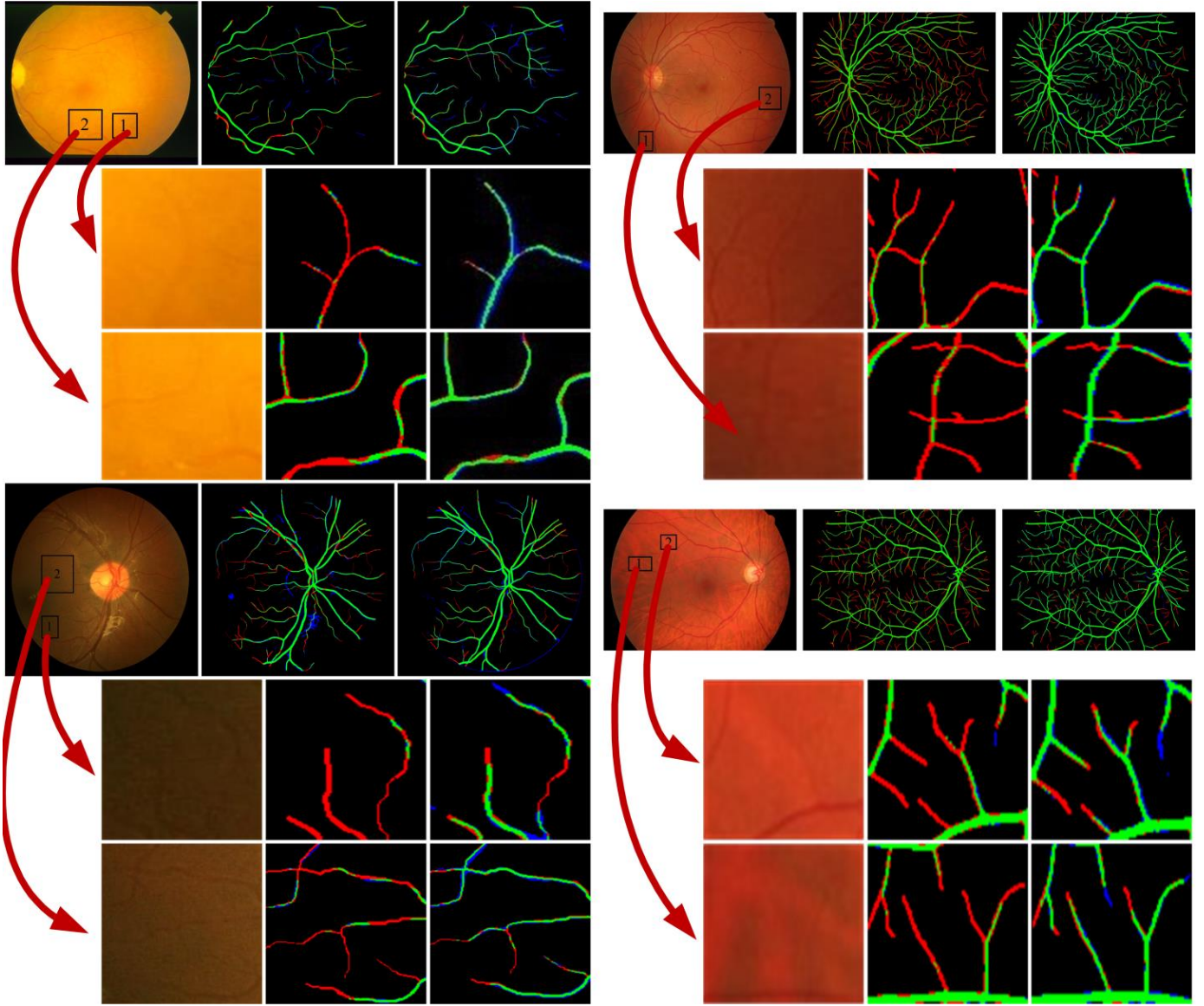

Fig. 6 Elusive vessel segmentation performance compared with Yan et al work [24].Examples comes from the test results on the four datasets. The first and fourth rows shows the global pixels distribution, and local detail segmentation maps are displayed in row 2, 3, 5, and 6. The images from column 1 to 3 are original fundus images, segmentation pixels statistic results with [24], and statistic results with the proposed method. Green indicates TP, blue represents FP and red means FN.

Table Ⅰ provides the quantitative performance and comparison calculated with our network and state-of-the-art methods. In the DRIVE dataset, *Se* ranks first, with a significant improvement of 0.0256 compared with the previous highest score in the work of Wu *et al.* [42]. The *Sp* shows a slight decrease of 0.0062 compared with that in Lupascu *et al.*'s method [22] whereas *Se* is 0.1566 higher. The results *AUC* is optimal, being at 0.9830. The *Acc* is close to the highest level (i.e., difference of 0.0018). Meanwhile, the *G* surpasses the second position over 0.0145, and our network occupies leading position in terms of *Pr* and *F1-Score*. It is worth noting that we compete with different state-of-the-art methods, and our network displays multi best result and strongest comprehensive capability.

In STARE dataset, the method of Liskowski and Krawiec [37] shows powerful effects in *Sp*, *Acc*, *AUC*, and other metrics. However, such method requires complex preprocessing steps including global contrast normalization, zero-phase component analysis, geometric transformations and gamma corrections. Our network needs no complex schedule and achieves 0.8812 in *Se* and 0.9283 in *G-Mean*. Furthermore, we still rank first in *Pr* and *F1-Score* with dramatic advantages. In conclusion, it is a matched rival for our method with [37] in STARE, outperforming other methods to a large degree, but our network is more efficient in the retinal vessel segmentation, as elaborated in the *E. Computation Efficiency*.

As for CHASEDB1 dataset, the results are shown in Table Ⅱ. *Se* achieves 0.8435, ranking first place with superiority of 0.0303. The *Sp* is only 0.0083 lower than the best score in Lyu *et al* method [39] while *Se* is 0.0557 higher. The *Acc* is 0.0034 lower than the best score. In the rest four metrics (*AUC*=0.9872, *G* =0.9083, *F1* =0.8218 and *Pr*=0.8013), our network outperforms all other algorithms. Especially in *Pr*, *G* and *F1*, significant enhancements in the performance are observed. Considering the performances on the three traditional datasets, the proposed network simultaneously works well. It always occupies the first position in *Se*, *G*, *Pr* and *F1*, and occasionally

TABLE I
COMPARATIVE PERFORMANCE OF THE PROPOSED NETWORK WITH EXISTING WORKS ON THE DRIVE AND STARE DATASETS, TOGETHER WITH THE ABLATION
EXPERIMENT RESULTS, INCLUDING NETWORK STRUCTURE ABLATION AND LOSS FUNCTION ABLATION
(W/O REPRESENTS WITHOUT, G INDICATES GENERATOR AND D MEANS DISCRIMINATOR)

| | DRIVE | | | | | | | STARE | | | | | | |
|---|---|---|---|---|---|---|---|---|---|---|---|---|---|---|
| Methods (year) | Se | Sp | Pr | Acc | AUC | G | F1 | Se | Sp | Pr | Acc | AUC | G | F1 |
| 2nd Human observer | 0.7760 | 0.9724 | 0.8066 | 0.9472 | - | 0.8686 | 0.7910 | 0.8951 | 0.9387 | 0.6424 | 0.9349 | - | 0.9166 | 0.7401 |
| **Unsupervised** | | | | | | | | | | | | | | |
| Fathi [38] (2013) | 0.7768 | 0.9759 | 0.7559 | **0.9581** | 0.9516 | 0.8706 | 0.7662 | 0.8061 | 0.9717 | 0.7027 | 0.9591 | 0.9680 | 0.8850 | 0.7508 |
| Zhao [8] (2015) | 0.7420 | 0.9820 | - | 0.9540 | 0.8620 | 0.8536 | - | 0.7800 | 0.9780 | - | 0.9560 | 0.9740 | 0.8734 | - |
| Azzopardi [3] (2015) | 0.7655 | 0.9704 | - | 0.9442 | 0.9614 | 0.8618 | - | 0.7716 | 0.9701 | - | 0.9497 | 0.9563 | 0.8651 | - |
| Zhang [40] (2016) | 0.7743 | 0.9725 | - | 0.9476 | 0.9636 | 0.8677 | - | 0.7791 | 0.9758 | - | 0.9554 | 0.9748 | 0.8719 | - |
| **Supervised** | | | | | | | | | | | | | | |
| Lupascu [20] (2010) | 0.6728 | **0.9874** | - | 0.9597 | 0.9561 | 0.8150 | - | - | - | - | - | - | - | - |
| Li [21] (2016) | 0.7569 | 0.9816 | - | 0.9527 | 0.9738 | 0.8619 | - | 0.7726 | 0.9844 | - | 0.9628 | 0.9879 | 0.8720 | - |
| Liskowski [37] (2016) | 0.7811 | 0.9807 | - | 0.9535 | 0.9790 | 0.8752 | - | 0.8554 | **0.9862** | - | 0.9729 | 0.9928 | 0.9184 | - |
| Orlando [23] (2017) | 0.7897 | 0.9684 | 0.7854 | - | - | 0.8744 | 0.7857 | 0.7680 | 0.9738 | 0.7740 | - | - | 0.8648 | 0.7644 |
| Wu [22] (2018) | 0.7844 | 0.9819 | - | 0.9567 | 0.9807 | 0.8776 | - | - | - | - | - | - | - | - |
| Yan [24] (2018) | 0.7653 | 0.9818 | - | 0.9542 | 0.9752 | 0.8668 | - | 0.7581 | 0.9846 | - | 0.9612 | 0.9801 | 0.8639 | - |
| Lyu [39] (2019) | 0.7940 | 0.9820 | - | 0.9579 | 0.9826 | 0.8830 | - | - | - | - | - | - | - | - |
| Wu [42] (2019) | 0.8038 | 0.9802 | | 0.9578 | 0.9821 | 0.8876 | - | - | - | - | - | - | - | - |
| Wang [43] (2019) | 0.7940 | 0.9816 | | 0.9567 | 0.9772 | 0.8828 | 0.8270 | - | - | - | - | - | - | - |
| **Proposed Method** | **0.8294** | 0.9812 | **0.8397** | 0.9563 | **0.9830** | **0.9021** | **0.8345** | 0.8812 | 0.9781 | **0.7952** | 0.9671 | 0.9863 | **0.9283** | **0.8359** |
| w/o AM | 0.8261 | 0.9793 | 0.8297 | 0.9540 | 0.9741 | 0.8939 | 0.8278 | 0.8652 | 0.9739 | 0.7870 | 0.9661 | 0.9814 | 0.9126 | 0.8296 |
| w/o MSFRB&AM | 0.7955 | 0.9823 | 0.8345 | 0.9521 | 0.9711 | 0.8839 | 0.8145 | 0.8364 | 0.9813 | 0.8083 | 0.9586 | 0.9790 | 0.9059 | 0.8221 |
| Only U-Net in G | 0.7886 | 0.9719 | 0.8043 | 0.9486 | 0.9705 | 0.8754 | 0.7963 | 0.7780 | 0.9835 | 0.7841 | 0.9633 | 0.9784 | 0.8747 | 0.7810 |
| Only U-Net in D | 0.6232 | 0.9792 | 0.7451 | 0.9333 | 0.9143 | 0.7811 | 0.6787 | 0.5904 | 0.9764 | 0.7640 | 0.9412 | 0.9408 | 0.7592 | 0.6660 |
| w/o GAN Loss | 0.7667 | 0.9685 | 0.7424 | 0.9319 | 0.9645 | 0.8617 | 0.7543 | 0.7572 | 0.9765 | 0.7704 | 0.9609 | 0.9752 | 0.8598 | 0.7637 |
| w/o MAE& BCE | 0.1349 | 0.9110 | 0.1747 | 0.8119 | 0.4383 | 0.3505 | 0.1522 | 0.2216 | 0.7967 | 0.1076 | 0.7374 | 0.3901 | 0.4201 | 0.1448 |
| w/o MAE | 0.8063 | 0.9782 | 0.8396 | 0.9575 | 0.9784 | 0.8881 | 0.8226 | 0.8544 | 0.9817 | 0.7848 | 0.9682 | 0.9868 | 0.9158 | 0.8181 |

TABLE II
COMPARATIVE PERFORMANCE OF THE PROPOSED NETWORK WITH EXISTING WORKS ON THE CHASEDB1 AND HRF DATASETS, TOGETHER WITH THE ABLATION
EXPERIMENT RESULTS, INCLUDING NETWORK STRUCTURE ABLATION AND LOSS FUNCTION ABLATION
(W/O REPRESENTS WITHOUT, G INDICATES GENERATOR AND D MEANS DISCRIMINATOR)

| | CHASEDB1 | | | | | | | HRF | | | | | | |
|---|---|---|---|---|---|---|---|---|---|---|---|---|---|---|
| Methods (year) | Se | Sp | Pr | Acc | AUC | G | F1 | Se | Sp | Pr | Acc | AUC | G | F1 |
| 2nd Human observer | 0.7760 | 0.9724 | 0.8066 | 0.9472 | - | 0.8686 | 0.7910 | - | - | - | - | - | - | - |
| **Unsupervised** | | | | | | | | | | | | | | |
| Azzopardi [3] (2015) | 0.7585 | 0.9587 | - | 0.9387 | 0.9487 | 0.8527 | - | - | - | - | - | - | - | - |
| Zhang [40] (2016) | 0.7626 | 0.9661 | | 0.9452 | 0.9606 | 0.8583 | - | 0.7978 | 0.9717 | | 0.9556 | 0.9608 | 0.8804 | - |
| **Supervised** | | | | | | | | | | | | | | |
| Li [21] (2016) | 0.7507 | 0.9793 | - | 0.9581 | 0.9716 | 0.8574 | - | - | - | - | - | - | - | - |
| Liskowski [37] (2016) | 0.7816 | 0.9836 | - | 0.9535 | 0.9823 | 0.8768 | - | - | - | - | - | - | - | - |
| Orlando [23] (2017) | 0.7277 | 0.9712 | 0.7438 | - | - | 0.8406 | 0.7332 | 0.7874 | 0.9584 | 0.6630 | - | - | 0.8687 | 0.7158 |
| Wu [22] (2018) | 0.7538 | 0.9847 | - | 0.9637 | 0.9825 | 0.8615 | - | - | - | - | - | - | - | - |
| Yan [24] (2018) | 0.7633 | 0.9809 | - | 0.9610 | 0.9781 | 0.8652 | - | 0.7881 | 0.9592 | 0.6647 | 0.9437 | - | 0.8694 | - |
| Lyu [39] (2019) | 0.7878 | **0.9865** | - | 0.9664 | 0.9865 | 0.8815 | - | - | - | - | - | - | - | - |
| Wu [42] (2019) | 0.8132 | 0.9814 | | 0.9661 | 0.9860 | 0.8876 | - | - | - | - | - | - | - | - |
| Wang [43] (2019) | 0.8074 | 0.9821 | | 0.9661 | 0.9812 | 0.8904 | 0.8037 | - | - | - | - | - | - | - |
| **Proposed Method** | **0.8435** | 0.9782 | **0.8013** | 0.9630 | **0.9872** | **0.9083** | **0.8218** | 0.8310 | 0.9730 | 0.8115 | 0.9559 | 0.9693 | 0.8992 | 0.8211 |
| w/o AM | 0.8392 | 0.9760 | 0.7966 | 0.9619 | 0.9834 | 0.9050 | 0.8173 | 0.8297 | 0.9702 | 0.8119 | 0.9542 | 0.9631 | 0.8972 | 0.8207 |
| w/o MSFRB&AM | 0.8258 | 0.9788 | 0.7617 | 0.9675 | 0.9765 | 0.8990 | 0.7924 | 0.8185 | 0.9757 | 0.7896 | 0.9503 | 0.9625 | 0.8890 | 0.8037 |
| Only U-Net in G | 0.7934 | 0.9706 | 0.7575 | 0.9625 | 0.9726 | 0.8775 | 0.7750 | 0.7763 | 0.9647 | 0.7609 | 0.9527 | 0.9620 | 0.8653 | 0.7685 |
| Only U-Net in D | 0.3650 | 0.9548 | 0.4404 | 0.9021 | 0.8092 | 0.5903 | 0.3991 | 0.2130 | 0.9635 | 0.3128 | 0.9388 | 0.8543 | 0.4530 | 0.2534 |
| w/o GAN Loss | 0.7755 | 0.9660 | 0.7531 | 0.9589 | 0.9654 | 0.8655 | 0.7641 | 0.7524 | 0.9639 | 0.7709 | 0.9483 | 0.9555 | 0.8516 | 0.7615 |
| w/o MAE&BCE | 0.1254 | 0.9021 | 0.1290 | 0.8248 | 0.5075 | 0.3363 | 0.1271 | 0.1844 | 0.9326 | 0.2235 | 0.7918 | 0.7147 | 0.4146 | 0.2020 |
| w/o MAE | 0.8387 | 0.9738 | 0.7803 | 0.9625 | 0.9841 | 0.9037 | 0.8084 | 0.8298 | 0.9719 | 0.7919 | 0.9534 | 0.9649 | 0.8980 | 0.8104 |

TABLE III
RESULTS OF THE CROSS-TRAINING EXPERIMENT D REPRESENTS DRIVE, S MEANS STARE, C INDICATES CHASEDB1, AND H IS HRF DATASET

| Test | Train | Method | Se | Sp | Acc | AUC |
|---|---|---|---|---|---|---|
| D | S | Li [21] | 0.7273 | 0.9810 | 0.9486 | **0.9677** |
| | | Yan [24] | 0.7292 | 0.9815 | 0.9494 | 0.9599 |
| | | **Proposed** | **0.7412** | **0.9830** | **0.9519** | 0.9643 |
| | C | Li [21] | 0.7307 | 0.9811 | **0.9484** | 0.9605 |
| | | **Proposed** | **0.7457** | **0.9865** | 0.9480 | **0.9661** |
| | H | Proposed | 0.8462 | 0.9640 | 0.9432 | 0.9671 |
| S | D | Li [21] | 0.7027 | 0.9828 | 0.9545 | 0.9671 |
| | | Yan [24] | 0.7211 | **0.9840** | 0.9569 | **0.9708** |
| | | **Proposed** | **0.8334** | 0.9764 | **0.9613** | 0.9718 |
| | C | Li [21] | 0.6944 | **0.9831** | 0.9536 | 0.9620 |
| | | **Proposed** | **0.8137** | 0.9765 | **0.9605** | **0.9785** |
| | H | Proposed | 0.8661 | 0.9599 | 0.9518 | 0.9664 |
| C | D | Li [24] | 0.7118 | **0.9791** | 0.9429 | 0.9628 |
| | | **Proposed** | **0.8134** | 0.9750 | **0.9547** | **0.9758** |
| | S | Li [24] | 0.7240 | **0.9768** | 0.9417 | 0.9553 |
| | | **Proposed** | **0.7950** | 0.9743 | **0.9485** | **0.9699** |
| | H | **Proposed** | 0.8364 | 0.9712 | 0.9582 | 0.9733 |
| H | D | Proposed | 0.8003 | 0.9779 | 0.9465 | 0.9541 |
| | S | Proposed | 0.8140 | 0.9785 | 0.9451 | 0.9579 |
| | C | Proposed | 0.8174 | 0.9710 | 0.9525 | 0.9509 |

achieves highest score in *AUC*.

In the challenging high resolution dataset HRF, the quantitative result with proposed network outperforms the state-of-the-art methods, as shown in Table II. Compared with second highest scores, there are markable enhancement in *Se*, *Pr*, *G*, *AUC*, and *F1*, respectively increased by 0.0332, 0.1468, 0.0188, 0.0085, and 0.1053. The *Sp* and *Acc* also take the first place with moderate improvement. The proposed method works well and maintains the outperforming performance as in the STARE, DRIVE, and CHASEDB1.

### B. Ablation Studies

For examining the benefits brought by SEGAN, MSFRB, and AM. We devise several ablation experiments to verify their functions. The whole results are contained at the bottle of Table II and Table I.

*1) Validation of SEGAN:* We retain the SEGAN structure to compare with two non-SEGAN networks, namely, "U-Net only in G" and "U-Net only in D". "U-Net only in G" includes a U-Net in G and a three-stage fully convolution network in D, while the other one swaps the network in G and D. The "U-Net only in D" works poorly which proves that the G must be a powerful network. Comparing the 'w/o MSFRB&AM' with "U-Net only in G", all metrics have been improved. The most attractive highlight includes improvements in *Se*, *Pr*, *G*, and *F1* with average values of 0.0349, 0.0218, 0.0212, and 0.0279 on DRIVE, STARE, CHASEDB1, and HRF datasets. In conclusion, the G structure needs to be powerful in multi-scale features learning to segment vessels well. The equilibrium D structure extracts multi-scale features to exert high pressure to G so that G could be trained better.

*2) Validation of MSFRB:* From the result between group 'w/o AM' and 'w/o MSFRB&AM'. It is obviously that MSFRB ma-

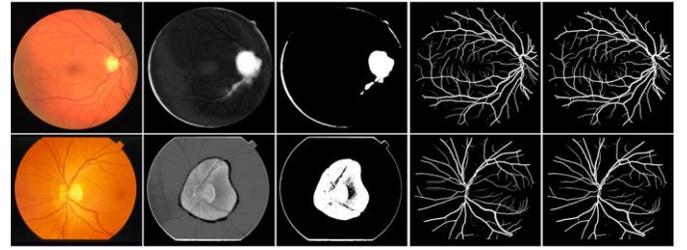

Fig. 7 Segmentation possibility map and segmented vessel of only GAN loss (column 2 to 3), and GAN loss together with BCE (column 4 to 5).

TABLE IV
COMPARISON ON THE COMPUTATION EFFICIENCY

| Method | Training time (h) | Computation time per image(s) |
|---|---|---|
| Liskowski [37] | 8 | 92 |
| Lyu [39] | **1** | 1.3 |
| Wu [22] | 16 | 10 |
| **Proposed** | 6 | **0.16** |

inly contributes to the *Se*, with improvements of 0.0306, 0.0288, 0.0134, and 0.0112 respectively on the four datasets, consistent with its theoretical function. By contrast, *Sp* slightly decreases in the four datasets. In other indicators, some fluctuations are observed, such as a declination of 0.0168 in *Pr* in DRIVE and a growth of 0.0349 in CHASEDB1. Although the MSFRB introduces a branch path in merging process, the structure requires limited extra trainable parameters and the added tensors are light, which will be elaborated in part *E. Computation Efficiency*.

*3) Validation of AM:* As seen in result, the AM raises the network overall performance, with more or less enhancement in each metric. Most metrics are ameliorated to the first rank. For instance, *Se* achieves top scores, figuring at 0.8294, 0.8812, 0.8435 and 0.8310 respectively on the four datasets, so are the metrics *G*, *F1* and *AUC*. It is noting that this part needs no extra trainable parameter which is fully self-motivated.

### C. Elusive Vessel Recognition

The detail information learning capability of the proposed symmetric network helps recognize the elusive vessels. As shown in Fig. 6, we compare the segmentation effect in more observable way. As the disturbance of lesion regions and low-contrast background, the original patches are extremely challenging to identify with the human eye. Compared the state-of-the-art method [24] with ours, the thick vessels are segmented well in two methods. However, the performances notably vary in the elusive vessels, especially in challenging situations. In our method, these challenging vessels are detected with higher probability, which are marked with green lines. The existing methods could find the thick vessels well, but the performance on elusive vessel recognition is limited. Intuitively, our method achieves good performance on the elusive vessels. Actually, this explains the reason why there is remarkable improvement in the quantitative metric *Se* and *Pr*, as much more elusive pixels are recognized with our method.

### D. Cross-training Evaluation

The practical extendibility of our network in diversity application is evaluated. We conduct a similar cross-training stra-

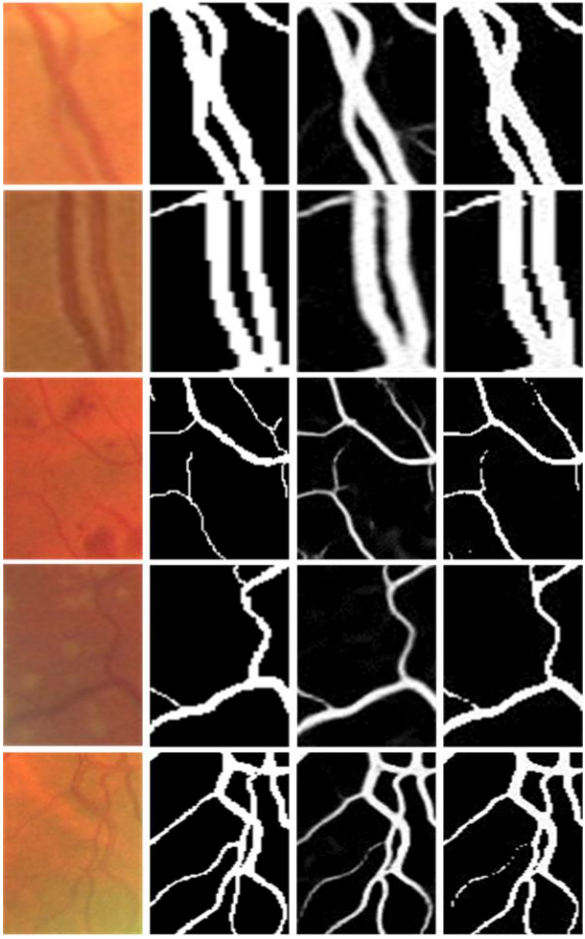

Fig. 8 Exemplar results on the challenging cases. The four columns are original patch, ground truth, segmentation possibility map, and binary segmentation map. The first two rows indicates the results of closely parallel vessels. The rows 3 and 4 are results when deal with images with lesions. The final row is the segmentation vessel in a background with gradient changing color.

### E. Computation Efficiency

The number of trainable parameters is about twice as much as one of U-net. The average training time of the proposed network for DRIVE dataset is 6 h with one NVIDIA Titan Xp GPU. After training, the processing time for an image only requires 0.16*s*. Firstly, the network could be trained end-to-end without complex preprocessing nor sub-processing. Second, the network only needs G to segment vessels from fundus image. Thus the network could drop approximate half weights after training, being light-weight and efficient (With our setting, there are about 8.6M trainable parameters in G, 10% heavier than traditional U-net structure). For quantitative comparison, we take a computation efficiency comparison with some methods, and the results are shown in Table Ⅳ.

## V. DISCUSSION

### A. Necessity of Optimized Loss Function

Extensive research has been conducted to evaluate the significance of combined loss function. We separately set several experiment groups, as shown in Table Ⅱ and Table Ⅰ. From the results, we learn that additional pixel classification loss functions are vital to highly enhance network performance because vessel segmentation belongs to the pixel classification task. The results with only GAN loss almost demonstrate no segmentation ability, as shown in Fig. 7. After combined with *BCE*, the results are extremely closed to the optimization level. The results indicate that supplemented segmentation loss functions are essential for granting a network optimal vessel pixel classification capability. When the GAN loss is removed, the overall performance evidently declines. In this case, the three components of the combined loss function are necessary in the proposed method.

### B. Performance on Challenging Cases

Compared with former researches, we have verified the proposed method's capability on overall vessel segmentation, based on Table Ⅰ and Table Ⅱ, especially for elusive retinal vessel segmentation, as shown in Fig. 6. Here we test the performance on three kinds of difficult situations. The first one is the segmentation task in the presence of lesions disturbance, and the second one is the task in dealing with closely parallel vessels when processing densely distributed vessels. The last one is the background color changing in some fundus images. As shown in Fig. 8, the proposed method distinguishes the closely parallel vessels, which avoids the wrongly mix. The vessels disturbed by lesions and background are also recognized well, which verify the capability of the proposed method in challenging cases.

### C. Improvement Room on the Specificity

The proposed network considerably enhances the performance on fundus vessel segmentation, especially in elusive vessels. However, *Sp* declines slightly compared with the highest score at present in the three of the four datasets ( $\Delta Sp = -0.0062$ in DRIVE, $\Delta Sp = -0.0081$ in STARE, $\Delta Sp = -0.0083$ in CHASEDB1, and $\Delta Sp = 0.0013$ in HRF), which show some non-vessel pixels are wrongly regarded as tegy to [24], training our network on one dataset and test on another one. Although a decline occurs in the performance, the results on cross-training experiment are still satisfactory and outperform other methods, as shown in Table Ⅲ.

In DRIVE test, the network trained on STARE yields high scores simultaneously in *Se*, *Sp* and *Acc* (i.e., 0.7412, 0.9830, and 0.9519 respectively). As for the model trained on CHASEDB1, all the indicators, except for Acc, rank first in DRIVE test. The model trained on HRF achieves good performance, especially with a high metric *Se*. In STARE test, the network trained on DRIVE achieves optima scores in *Se*, *Acc*, and *G*. Although *Sp* decays by 0.0076, the *Se* increases by 0.1123, indicating high recognition ratio of vessel. The good performance in *Acc* (0.9605) and *AUC* (0.9785) imply the comprehensive capability although there is no comparison method. Considering the test on CHASEDB1, the three networks trained on DRIVE, STARE, and HRF show markable improvement in *Se*, exceeding Li *et al.* work [21] of 0.1016 and 0.0710. Additionally, *Acc* and *AUC* have been increased approximately by 0.1, showing a strong general capability. For test on HRF, although there is no comparison group, the four metrics show satisfactory segmentation performance.

vessels. Specifically, the decay in *Sp* mainly originates from MSFRB (-0.0040 and -0.0074 in DRIVE and STARE, respectively) and partly from segmentation loss function. MSFRB enhance the search on local detail information and high resolution representation capability, while some indistinguishable details are classified as vessel pixels, which declines the *sp*.

Although additional *FP* pixels exist, the branch structure and shape consisting of these pixels are approximately coincidence with *TP* pixels (i.e., the location and distribution of *FP* pixels usually follow the *TP* pixels, instead of random component distribution, as shown in Fig. 6). In this case, the disturbance of *FP* pixels brought by the presented network is markedly weakened. The future work could focus on restraining the *FP* pixels production based on implicit information extracted from fundus images and prior knowledge of vessel structure.

## VI. CONCLUSION

In this work, we have presented a method to strengthening retinal vessel segmentation capability, especially on elusive vessels in low-contrast background and lesion regions. We refine the vessel detail information extraction by the proposed SEGAN and MSFRB, which highlight the multi-scale features learning and preserve the high resolution features with concise merging structure. The AM is also employed to distribute more attention on the discriminative feature, which improve overall performance. Through segmentation experiment, cross-training experiment, and ablation study, we verify the satisfactory vessel segmentation capability of our method. With it, much elusive vessel pixels are correctly classified, which is confirmed by the metrics *Sp*, *Pr*, *G*, *F1* and binary segmentation map. The challenges in vessel segmentation are solved well. The high computation efficiency provides considerable potential in clinical application. Meanwhile, the detail information learning capability of this method could be employed in other biomedical issues, not only in retinal vessel.